# Electronic Transport in Monolayer Graphene with Extreme Physical Deformation: *ab Initio* Density Functional Calculation

Haiyuan Gao[1], Yang Xu[1,a)], Meijiao Li[1], Zhendong Guo[1], Hongshen Chen[1], Zhonghe Jin[1], and Bin Yu[2,b)]

[1]Department of Information Science and Electronic Engineering, Zhejiang University, Hangzhou 310027, China
2 College of Nanoscale Science and Engineering, State University of New York, Albany 12203, USA

**Abstract:** Electronic transport properties of monolayer graphene with extreme physical bending up to 90° angle are studied using *ab Initio* first-principle calculations. The importance of key structural parameters including step height, curvature radius and bending angle are discussed how they modify the transport properties of the deformed graphene sheet comparing to the corresponding flat ones. The local density of state reveals that energy state modification caused by the physical bending is highly localized. It is observed that the transport properties of bent graphene with a wide range of geometrical configurations are insensitive to the structural deformation in the low-energy transmission spectra, even in the extreme case of bending. The results support that graphene, with its superb electromechanical robustness, could serve as a viable material platform in a spectrum of applications such as photovoltaics, flexible electronics, OLED, and 3D electronic chips.

## 1. Introduction

Graphene, a 2D carbon allotrope with outstanding electrical [1] and mechanical [2] properties, has been exploited for potential applications such as transparent electrodes, interconnects, and NEMS oscillators [3-5]. The physical deformation of 2D graphene sheet has been observed due to specific fabrication process or surface topography of the supporting substrate. While minor deformation was reported in graphene on top of rough surface of dielectric materials (e.g. $Al_2O_3$ [6] and $SiO_2$ [7]), graphene sheets with extreme physical bending were reported using pre-fabricated structural template such as oxide trench [8]. Key material and electrical properties of graphene lying on corrugated or stepped surface need to be examined. Physically deformed graphene nanoribbons with a 14Å step in height were studied recently [9]. However, little attention has been devoted to how the varying geometry configuration influences the electronic transport properties of step-shaped graphene. In this work, we investigate the key material and transport properties of dramatically deformed (step-shaped) graphene sheets with varying geometry configuration (i.e., step height, curvature radius, and bending angle). Previous Raman spectroscopy analysis suggests negligible strain of 0.025% [8], so we focus on step-shaped graphene in absence of strain. Our work can provide guidance to implementation and engineering of "3D graphene" systems in potential technology applications.

## 2. Simulation

Motivated by our previous experimental research of 3D stepped graphene [8], we explore the physically deformed graphene in a more systematical approach. The simulated structure is schematically shown in Fig. 1(a) in which structural parameters are illustrated, such as step height (*H*), curvature radius (*CR*), and bending angle (*θ*). Periodic boundary conditions are imposed in the width direction (x-axis) and semi-infinite electrodes are used in the transport direction (z-axis) to describe two-dimensional devices. Because graphene nanoribbons with armchair edge-shape has larger band-gap than that of zigzag edge-shape [10], we use the channel region along armchair edge-shape to explore the maximum influences of deformation on electronic properties. We use the number of dimer line ($N_a$) in the unit cell of the channel region, representing the channel length. $N_a$ divides the band-gap of the channel region into three families and $N_a = 3p + 1$ ($p$ is integer) is the maximum band-gap family [10]. To examine the maximum influences of deformation, we choose $N_a = 40$ in the maximum band-gap family, which is equivalent to 4.8 nm long flat (non-deformed) graphene. The flat graphene is bent into various structural shapes for the study of electronic transport dependence on deformation. These shapes can be obtained through pre-fabricated structural template as shown in Fig. 1(b). A bias voltage $V_b$ is applied between the semi-infinite electrodes in two ends. The bending arcs of graphene is modeled as a partial carbon nanotube (CNT) ring [11]. *CR* is calculated from the chiral vector of the corresponding CNT.

Electronic properties are obtained using SIESTA packages [12], in which numerical atomic orbitals are used as basis sets and Troullier-Martin type norm-conserving pseudo-potentials. The exchange-correlation function of the local density approximation (LDA) is represented by the Ceperley-Alder approximation [13]. Total 200 Ry mesh cutoff is chosen. The convergence criterion for the density matrix is taken as $1\times10^{-3}$. After testing and comparing various converged *k*-meshes and basis sizes, we choose a double ζ-plus-polarization basis set with 500×1×1 for LDOS calculations and a single ζ basis set with 100×1×1 for transport calculations. Interaction between adjacent graphene layers is hindered by a large spacing of 8Å.

Transmission spectra and current-voltage (I-V) characteristics are investigated by the Nonequilibrium Green's Function (NEGF) approach. It is based on the Density Function Theory (DFT) implemented in the TranSIESTA module (within the SIESTA package) [14]. The current through the electrodes is calculated using the Landauer-Buttiker formula [15]

$$I(V_b) = G_0 \int_{\mu_L}^{\mu_R} T(E,V_b)[f_L(E,\mu_R) - f_R(E,\mu_L)]dE , \qquad (1)$$

where $G_0 = 2(e^2/h)$ is the unit of quantum conductance and $h$ is the Planck's constant. $f_L$ and $f_R$ are the Fermi–Dirac distribution functions and $\mu_L$, $\mu_R$ are the chemical potentials of the left and right electrodes, respectively. $T(E,V_b)$ is the transmission probability of electrons (with an energy $E$) flowing through the device under the bias potential $V_b$. Temperature is 0 K in our simulation.

## 3. Results and discussion

When calculating electrostatic properties, we focus on the localization of energy states in the channel region of bent graphene, excluding the semi-infinite electrodes. Due to the finite length (4.8 nm) of the channel region, we build a device with periodic boundary conditions in the z-axis to simulate two-dimensional graphene, as shown in Fig. 2(a). The local density of states (LDOS) at sampling atoms are shown in Fig. 2(b). Difference in LDOS between flat and bent graphene is significant only at the atom "Ⅳ" in the bent arc. For the sampling atoms located further away from the arc (e.g. atom "Ⅰ"), the LDOS of flat and bent graphene is almost identical. These results indicate that the electronic states induced by physical deformation are highly localized, only in the bent arc. Comparing all of the four representative sampling atoms in Fig. 2(b), the LDOS modification due to bending appears at high energy region rather than around the Fermi level.

To further investigate the transport mechanism, transmission spectra $T(E, V_b)$ at various $V_b$ are plotted in Fig. 3. At $V_b = 0.0$ V, near-linear transmission spectra are found. Transmission of bent graphene is close to that of flat graphene near the Fermi level. The differences of transmission spectra between bent and flat graphene in high energy regions are consistent with the LDOS in Fig. 2(b). These results are attributed to the fact that deformation induces no significant change in the low-energy π bond network, as π bonds are not sensitive to bond angles. However, the σ bond network, which has energy far from the Fermi energy, is modified via changing the bond angle of adjacent bonds. With increasing $V_b$, the transmission spectra near the Fermi level are filled with broader channels. For example, at $V_b = 0.6$ V, the transmission of bent graphene is smaller than that of flat graphene at the Fermi energy. This is due to the broken symmetry of π bond network in bent graphene, which makes the effective transporting channels less than those in flat graphene [8]. By comparing transmission around the Fermi level, we find dependency of transmission on structural parameters is largest for curvature radius, and smallest for bending angle.

I-V characteristics of bent graphene are shown in Fig. 4, displaying no significant difference from those of flat graphene at low $V_b$. The small slope of I-V curves at low $V_b$ is attributed to the short channel region we simulated. When the bias voltage is larger than 0.3 V, linearity of I-V curve is "recovered". At high $V_b$, the current of bent graphene is smaller than that of flat graphene. This is because the transmission coefficient of bent graphene in low-energy region, which contributes most of the transport current, is smaller than that of flat graphene. As shown in Fig. 4(b), we observe curvature radius is the most determining parameter on I-V characteristics. With larger CR, the current of bent graphene is closer to that of flat graphene. These are due to that extreme small CR significantly changed the symmetry of graphene π bond network and reduce the transport channels. We also find the step height influences the current, as shown in Fig. 4(a). With increasing H, the current of bent graphene is closer to that of flat graphene. Theses are attributed to that large height represents long flat section between the two bent arcs, reducing the coupling between the arcs and making bent graphene close to the flat one. In Fig. 4(c), we observe no significant dependence of current on the bending angles, which means that we can ignore the slope of the beneath step shape in fabrication. Note that transport properties are insensitive to deformation within 0.085 V/nm electric field in the transport direction (equivalent to $V_b = 0.3$ V). Within this condition, graphene is suitable for devices with surface topography and flexible electronic applications.

## 4. Summary

In conclusion, the electronic transport properties of bent graphene sheet with various physically deformed configurations are compared. It is observed that the transmission spectra of the step-shaped graphene have negligible modification from its flat counterpart in low-energy region as a result of the insensitivity of transmission spectra from π-electron bond network on deformation. At high bias voltage, curvature radius are the most determining parameters on transport properties of bent graphene, then the step height in the middle, and the least important parameter is the bending angle. Transport properties of graphene are insensitive to deformation within certain electric field in the transport direction, making graphene suitable for devices with surface topography or flexible electronics applications.
.

**Acknowledgement:**

The authors would like to thank Prof. Jack Luo and Prof. Wei Ji for helpful discussion. This work is supported by the National Science Foundation of China (Grant No. 61006077) and National Basic Research Program of China (Grant No. 2006CB300405). Prof. Yang Xu is also supported by the Excellent Young Faculty Awards Program (Zijin Plan) at Zhejiang University and Specialized Research Fund for the Doctoral Program of Higher Education (SRFDP with Grant No. 20100101120045).

**References**

[1]  A. K. Geim, and K. S. Novoselov, Nat. Mater. **6**, 183 (2007).

[2]  C. Lee, X. D. Wei, J. W. Kysar, and J. Hone, Science **321**, 385 (2008).

[3]  Y. P. Chen, and Q. K. Yu, Nature Nanotech. **5**, 559 (2010).

[4]  Y. M. Lin, C. Dimitrakopoulos, K. A. Jenkins, D. B. Farmer, H. Y. Chiu, A. Grill, and P. Avouris, Science **327**, 662 (2010).

[5]   K. S. Kim, Y. Zhao, H. Jang, S. Y. Lee, J. M. Kim, J. H. Ahn, P. Kim, J. Y. Choi, and B. H. Hong, Nature **457**, 706 (2009).

[6]   L. Liao, J. W. Bai, Y. Q. Qu, Y. Huang, and X. F. Duan, Nanotechnology **21**, 015705 (2010).

[7]   M. Ishigami, J. H. Chen, W. G. Cullen, M. S. Fuhrer, and E. D. Williams, Nano Lett. **7**, 1643 (2007).

[8]   B. D. Briggs, B. Nagabhirava, G. Rao, R. Geer, H. Y. Gao, Y. Xu, and B. Yu, Appl. Phys. Lett. **97**, 223102 (2010).

[9]   Z. Z. Yu, L. Z. Sun, C. X. Zhang, and J. X. Zhong, Appl. Phys. Lett. **96**, 173101 (2010).

[10]  Y. W. Son, M. L. Cohen, and S. G. Louie, Phys. Rev. Lett. **97**, 216803 (2006).

[11]  A. Javey, J. Guo, Q. Wang, M. Lundstrom, and H. J. Dai, Nature **424**, 654 (2003).

[12]  J. M. Soler, E. Artacho, J. D. Gale, A. Garcia, J. Junquera, P. Ordejon, and D. Sanchez-Portal, J. Phys.: Condens. Matter **14**, 2745 (2002).

[13]  J. P. Perdew, and A. Zunger, Phys. Rev. B **23**, 5048 (1981).

[14]  M. Brandbyge, J. L. Mozos, P. Ordejon, J. Taylor, and K. Stokbro, Phys. Rev. B **65**, 165401 (2002).

[15]  J. Maassen, W. Ji, and H. Guo, Appl. Phys. Lett. **97**, 142105 (2010).

a) E-mail: yangxu-isee@zju.edu.cn.
b) E-mail: BYu@uamail.albany.edu

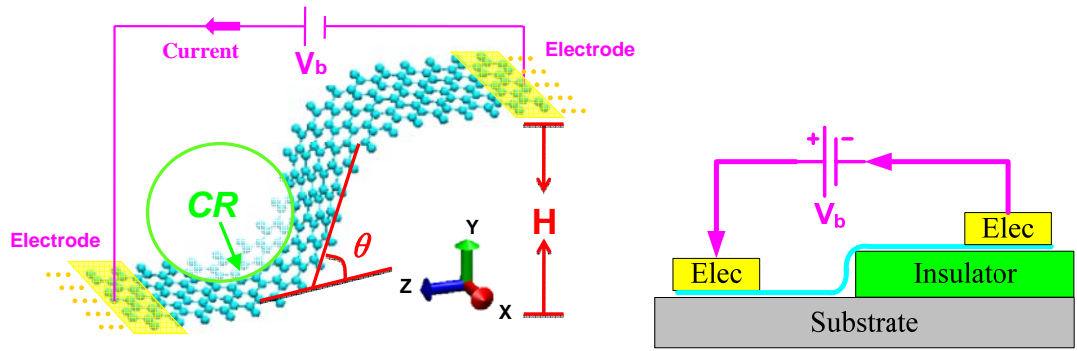

Fig. 1. (a) Schematic of step-shaped graphene with key structural parameters indicated. Blue balls represent carbon atoms. Periodic boundary conditions are imposed in the x-axis direction. Semi-infinite electrodes are set in the two ends to calculate transport properties. The current transport is along the z-axis direction. (b) Illustration of the step-shaped graphene laying on a stepped surface. The blue curve represents monolayer graphene.

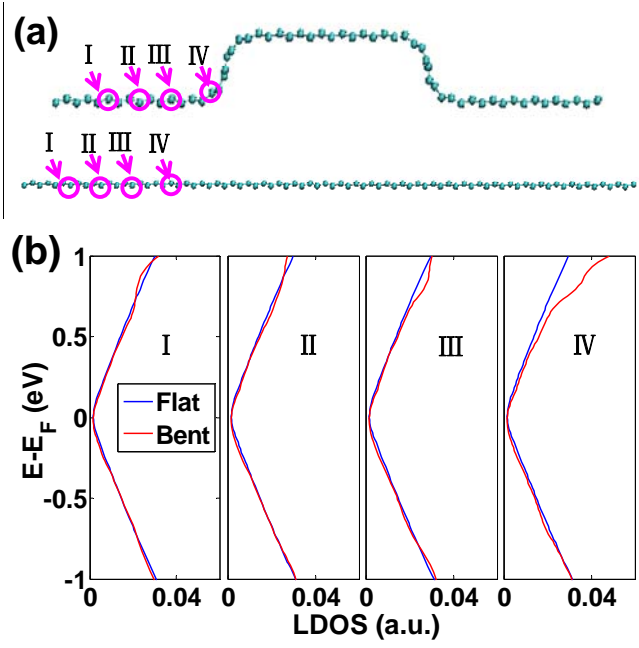

Fig. 2. (a) Illustration of the sampling atoms in LDOS calculation. (b) LDOS of bent and flat graphene at some sampling atoms. The bent graphene device has a step height of 0.78 nm, a curvature radius of 0.40 nm, and a bending angle of 90°.

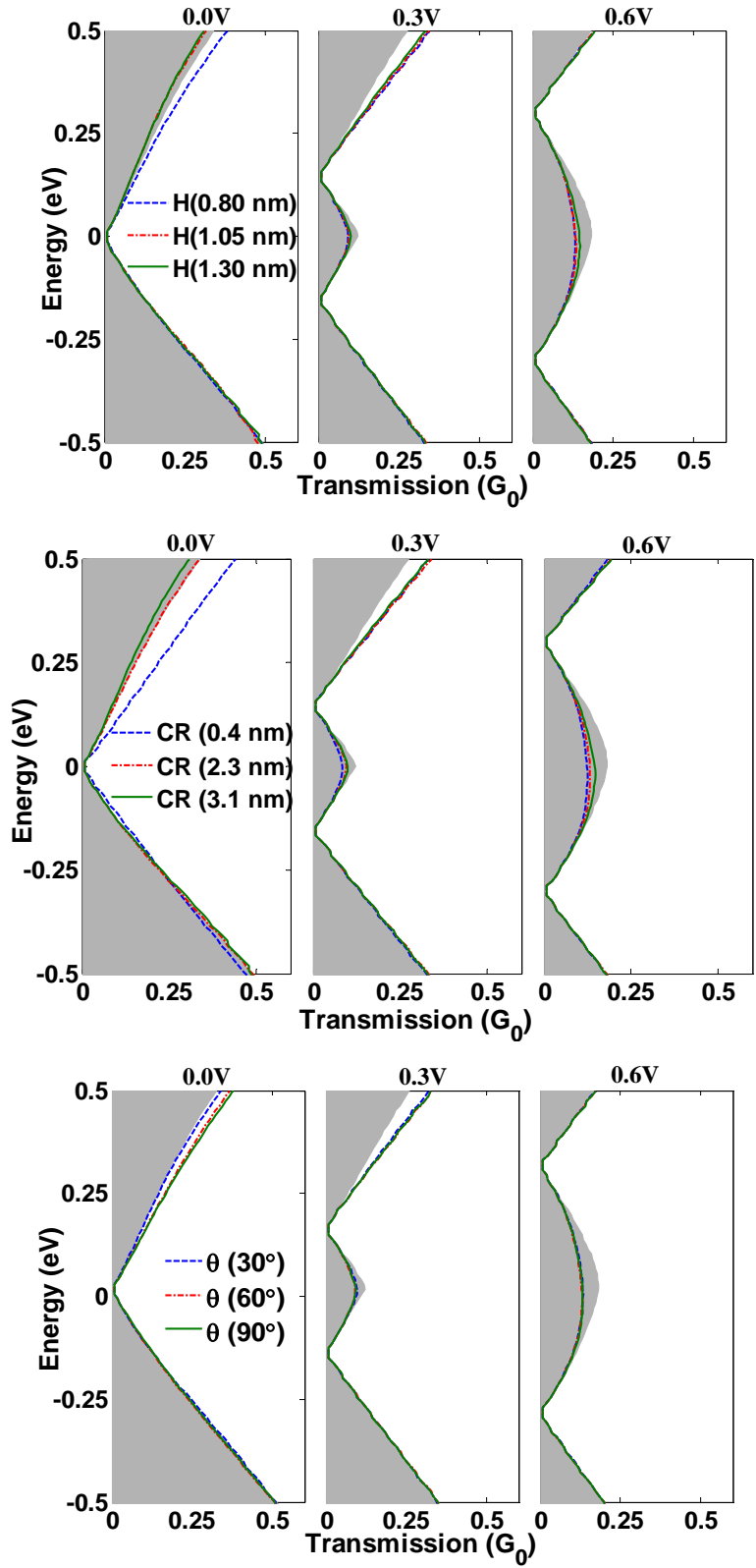

Fig.3 (a) Transmission spectrum of bent graphene with different step height (a), curvature radius (b), and bending angle (c) with bias voltage from 0.0 V to 0.6 V, as compared with that of flat graphene (the shaped area). CR is 3.1 nm and angle is 30° in (a). H is 1.3 nm and angle is 30° in (b). H is 2.3 nm and CR is 1.6 nm in (c).

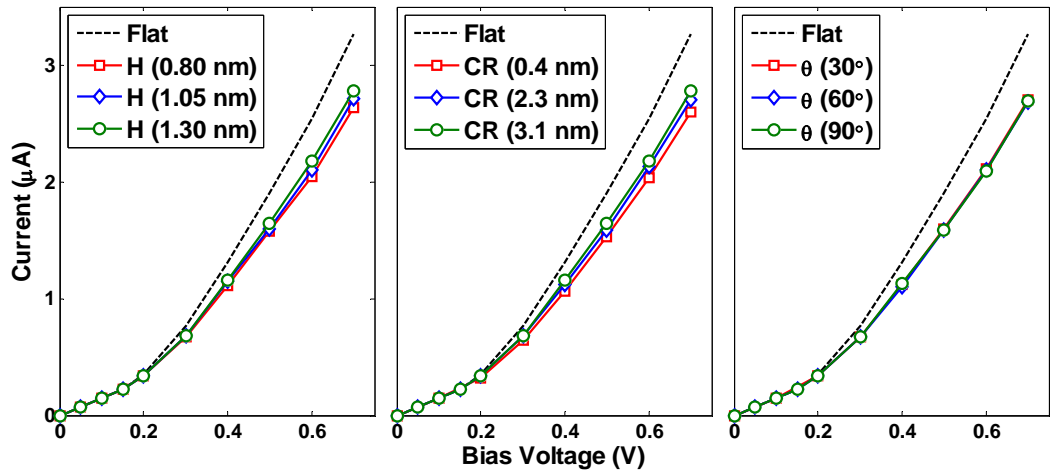

Fig.4: I-V characteristics of bent graphene with different step height (a), curvature radius (b), and bending angle (c). CR is 3.1 nm and angle is 30° in (a). H is 1.3 nm and angle is 30° in (b). H is 2.3 nm and CR is 1.6 nm in (c). The current of bent graphene is close to that of flat graphene at small bias voltage (while smaller at large bias voltage).